\documentclass[aip,apl,amsmath,amssymb,reprint]{revtex4-1}

\usepackage{graphicx}
\usepackage{dcolumn}
\usepackage{bm}
\usepackage{threeparttable}
\usepackage{amsmath}
\usepackage[english]{babel}
\usepackage{blindtext}
\usepackage{siunitx}
\usepackage{textgreek}
\usepackage{booktabs}
\usepackage{natbib}
\usepackage{multirow}

\begin{document}

\title{Effects of Interstitial Oxygen and Carbon on Niobium Superconducting Cavities}

\author{P. N. Koufalis}
\email{pnk9@cornell.edu}
\thanks{This work is supported by NSF Awards PHY-1416318 and PHY-1549132, the Center for Bright Beams.}
\author{D. L. Hall}
\author{M. Liepe}
\author{J. T. Maniscalco}
\affiliation{Cornell Laboratory for Accelerator-Based Sciences and Education (CLASSE), Cornell University, Ithaca, NY 14853}

\date{\today}

\begin{abstract}
We present results on the effects of interstitial oxygen and carbon on a bulk-niobium superconducting radio-frequency cavity. Previous experiments have shown that high-temperature ($\sim$800 $^\circ$C) nitrogen-doping plays the dominant role in the reduction of the electronic mean free path in the RF penetration layer of niobium that leads to a decrease in microwave surface resistance and a suppression the temperature-dependent component of the surface resistance with increasing accelerating gradient. In this work, we show that oxygen and carbon-doping has very similar effects on cavity performance, demonstrating that these effects are not unique to nitrogen. The preparation method used in the introduction of interstitial oxygen and carbon has the advantage that it is done at lower temperatures than that of high-temperature nitrogen-doping and does not require post-treatment electro-polishing.
\end{abstract}

\maketitle

\section{Introduction}
Decreasing the surface resistance of niobium superconducting radio-frequency cavities is of paramount importance for reducing cryogenic losses and operating costs of future accelerators. Previous experiments have shown that high-temperature ($\sim$800 $^\circ$C) nitrogen-doping of niobium cavities has two major advantages over traditionally prepared niobium cavities\cite{G13,G16}. The first advantage is an overall decrease of the microwave surface resistance, $R_S$, by the lowering of the electron mean free path in the RF penetration layer to values near the optimal value for minimum surface resistance. The second, more interesting effect is the suppression of the temperature-dependent component of the surface resistance, $R_{BCS}$, with increasing fields leading to quality factors $Q_0$ of up to $4\times10^{10}$ at $T = 2.0$ K\cite{G13}. This suppression of $R_{\text{BCS}}$ was first discovered at Jefferson Lab but was apparently caused by titanium impurities\cite{D13}.  Recent theoretical work offers a promising insight into the physical mechanism that drives these effects, but as yet there is not a strong consensus on the matter\cite{G14,M16}.

Experiments on niobium cavities first completed at Fermilab, in which the cavities were treated in a low-temperature nitrogen `atmosphere' ($T\leq200$ $^\circ$ C), exhibited the same behavior as the high-temperature nitrogen-doped cavities ($T \geq 600 ^\circ$C) and appeared to imply significant nitrogen diffusion into the niobium in quantities sufficient enough to reproduce this behavior\cite{TTC}. However, the results presented in this letter suggest that this is not the case. We demonstrate that it is the introduction of oxygen and carbon as interstitial impurities in the niobium lattice at concentration levels similar to that of the high-temperature nitrogen-doped cavities that leads to these behaviors.

\section{High Temperature Nitrogen Doping}
In the high-temperature nitrogen-doping process the niobium cavity is treated in a low-pressure ($P\approx40$ mTorr) nitrogen atmosphere at temperatures ranging from 600-1000 $^{\circ}$C typically followed by an annealing step in ultra-high vacuum \cite{G13,G16}. During the doping step, nitrides form on the surface of the cavity allowing a net flow of nitrogen from the nitride layer into the bulk lattice as an interstitial\cite{CR80,G16}. Following treatment, the cavity is electro-polished (EP) to remove the lossy nitride layer\cite{G13,G16}. We will see that oxygen and carbon-doped cavities do not require post-doping EP with the treatment method used.

\section{Low Temperature Treatment}
We heat treated a TESLA-shaped\cite{A00} 1.3 GHz niobium cavity and a niobium sample together with a three step process. The first step was done at 800 $^\circ$C in ultra-high vacuum for 10 hr. The second step was done at 160 $^\circ$C in a continuously flowing nitrogen `atmosphere' for 48 hr. The last and final step was done at 160  $^\circ$C in ultra-high vacuum for 168 hr. The steps were completed sequentially without removing the cavity and sample from the furnace.

\section{RF Performance}
\begin{figure}[h]
\centering{
\includegraphics[width=0.95\columnwidth]{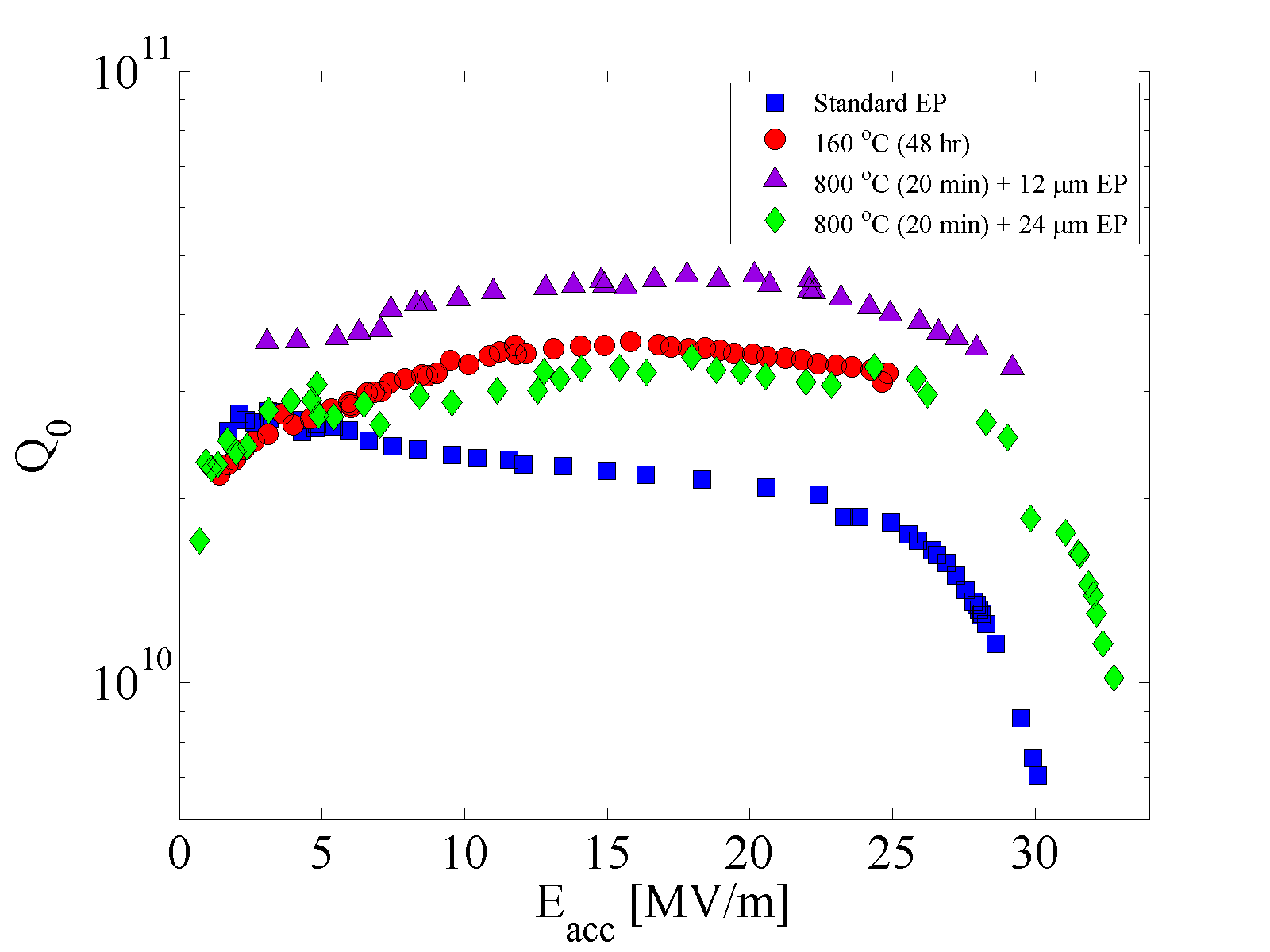}
\caption{RF cavity performance at $T$ = 2.0 K for two (800 $^\circ$C)nitrogen-doped cavities, the 160 $^\circ$C cavity, and a standard-treatment bulk-niobium cavity. The nitrogen doped cavities were doped at 800 $^\circ$C for 20 mins in a nitrogen atmosphere, annealed at 800 $^\circ$C in ultra-high vacuum for 30 mins, followed by an electro-polish. The standard treatment refers to a bulk electro-polish of a niobium cavity followed by high-pressure rinsing in ultra-high purity water and RF testing.}
\label{fig:three}}

\end{figure}

The RF performance of the 160 $^\circ$C cavity was comparable to that of the high-temperature nitrogen-doped cavities shown in Fig.\ \ref{fig:three}. It displayed the so-called `anti-$Q$-slope' with increasing accelerating gradient, $E_{\text{acc}}$, reaching a maximum $Q_0 = 3.6\times10^{10}$ at 2.0 K and 16 MV/m. 

The nitrogen-doped cavities were treated at 800 $^\circ$C for 20 min in a 40 mTorr nitrogen atmosphere followed by a 30 min anneal at 800 $^\circ$C in ultra-high vacuum. The concentration of nitrogen in the RF layer is controlled by the amount of post-treatment EP. It is important to note here that the 160 $^\circ$C cavity received no post-treatment EP. The 12 $\mu$m EP nitrogen-doped cavity achieves quality factors higher than that of the 24 $\mu$m EP nitrogen-doped cavity as its electronic mean free path, $\ell = 34$ nm, is closer to the mean free path ($\ell_{\text{min}} \approx 20$ nm) that minimizes the temperature-dependent component of the surface resistance, $R_{BCS}$, for a given accelerating field. The 160 $^\circ$C cavity, with a measured $\ell = 7.5$ nm, is considered to be heavily-doped. The 24 $\mu$m EP nitrogen-doped cavity with $\ell = 47$ nm behaves very similarly as it is nearly directly opposite of the minimum in the $R_{BCS}$ vs.\ $\ell$ curve.

For comparison, the RF performance of a standard bulk niobium cavity is also shown. It received an EP, 900 $^\circ$C degassing bake in ultra-high vacuum for 3 hr, and a high pressure rinse in de-ionized water before RF testing. The ubiquitous medium-field and high-field $Q$-slopes are evident \cite{H08}. 

\begin{figure}[h]
\centering{
\includegraphics[width=0.95\columnwidth]{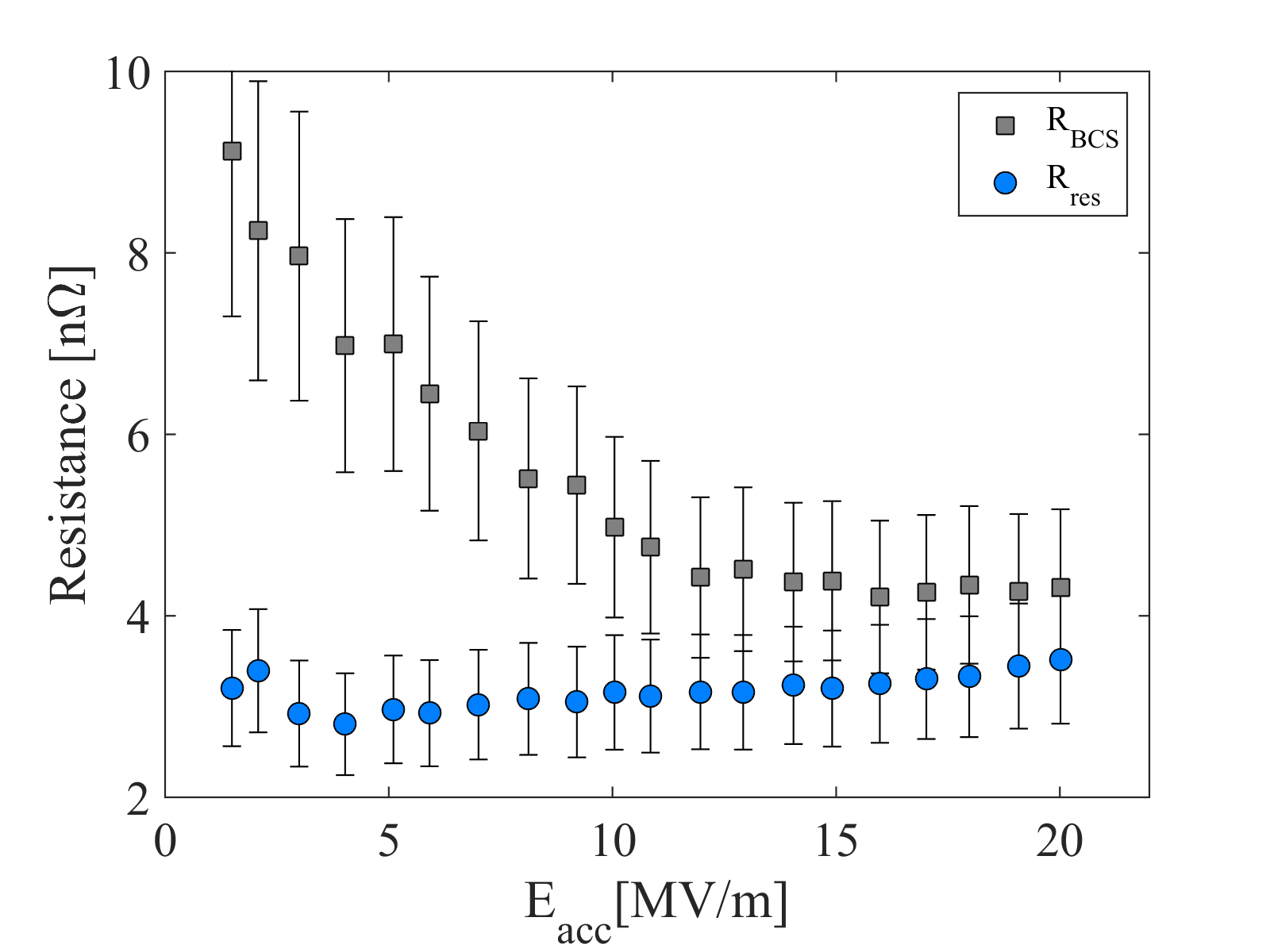}
\caption{Deconvolution of the temperature independent residual resistance, $R_0$, and the temperature-dependent BCS resistance, $R_{\text{BCS}}$, as a function of the accelerating gradient, $E_{\text{acc}}$, at 2.0 K.}
\label{fig:four}}
\end{figure}

Figure \ref{fig:four} shows a deconvolution of $R_{\text{BCS}}$ and $R_0$ as a function of accelerating gradient for the 160 $^\circ$C cavity. The residual resistance is relatively constant whereas $R_{\text{BCS}}$ decreases with increasing gradient demonstrating the effect of interstitial carbon and oxygen on the temperature-dependent $R_{\text{BCS}}$.

We used the SRIMP code developed by Halbritter\cite{H70} and translated into MATLAB by Valles\cite{V13} to extract the electron mean free path, $\ell$, quasi-particle energy gap, $\Delta(0)/k_BT_c$, and temperature-independent residual resistance, $R_0$, from data collected during RF testing of the 160 $^\circ$C cavity.  The mean free path is obtained from a BCS fit of measurements of the penetration depth, $\lambda$, vs.\ $T$ and was found to be 7 nm. This implies a significant amount interstitial impurities in the RF penetration layer. The energy gap and residual resistance are obtained from a BCS fit of the surface resistance as a function of $T$. Figure \ref{fig:two} shows the fits of both measurements and the relevant extracted material parameters.

\section{Secondary Ion Mass Spectroscopy}
\begin{figure}[h]
\centering{
\includegraphics[width=0.95\columnwidth]{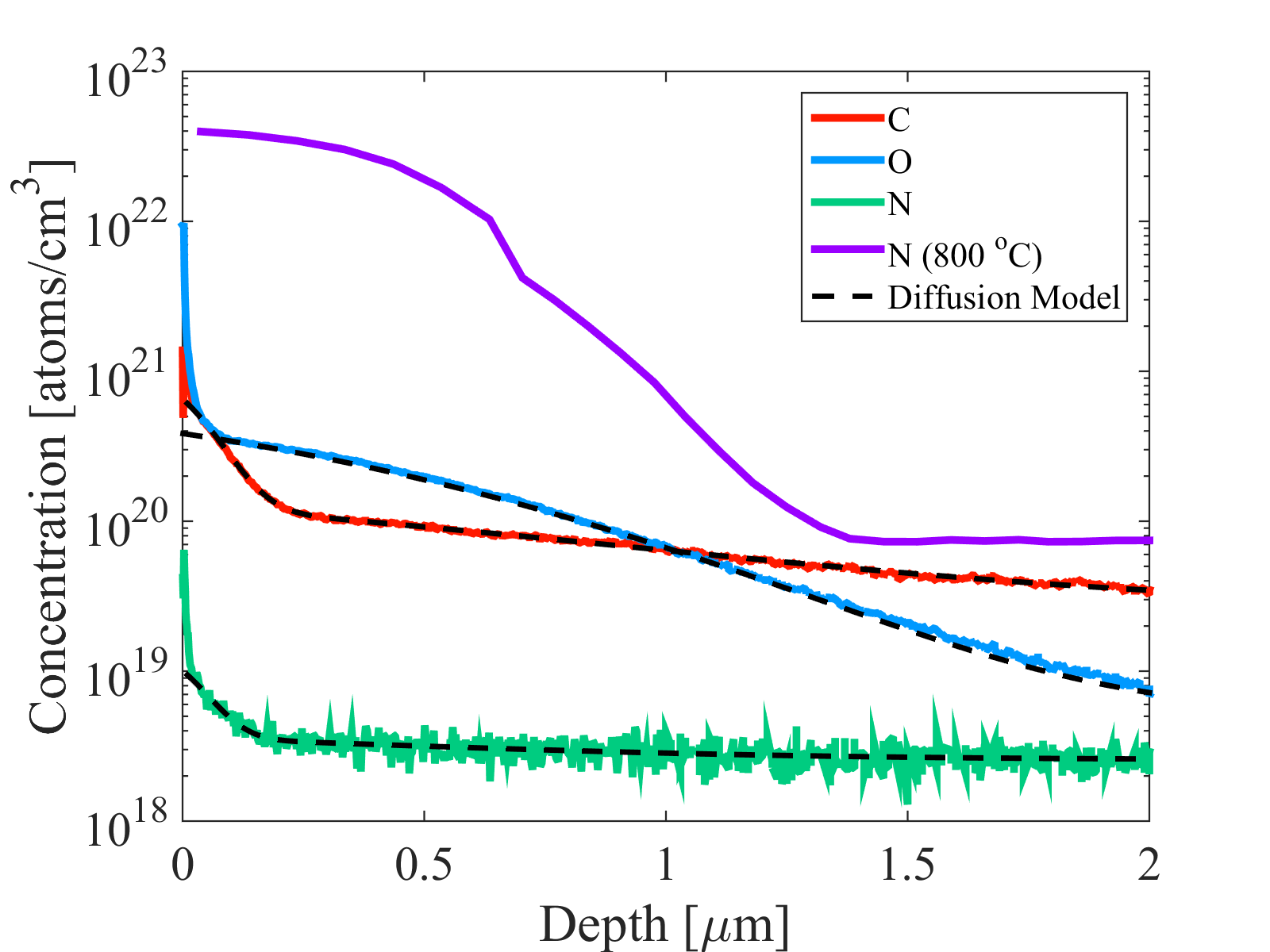}
\caption{Concentration profiles for C, O, and N as a functions of depth in a niobium sample treated, chronologically, at 800 $^\circ$C for 10 hr in ultra-high vacuum, 160 $^\circ$C for 48 hr in a continuously flowing nitrogen atmosphere with trace amounts of O$_2$ and H$_2$O (i.e. $\le$ 5 ppm), and 160 $^\circ$C for 168 hr in ultra-high vacuum. For comparison, a nitrogen concentration profile for a sample doped at 800$^\circ$C is shown.}
\label{fig:one}}
\end{figure}

Figure \ref{fig:one} shows the concentration profiles measured with Secondary Ion Mass Spectroscopy (SIMS) for C, O, and N in a niobium sample that received the 160 $^\circ$C treatment as well as a sample treated with a high-temperature (800 $^\circ$C) nitrogen-dope. Not surprisingly, there is a distinct difference in the nitrogen concentration of the 160 $^\circ$C sample compared to that of the high-temperature nitrogen-doped sample. The nitride layer of the nitrogen-doped sample ends at a depth of approximately 1.5 $\mu$m and is etched off prior to RF testing leaving an interstitial nitrogen concentration on the order of $\sim$7$\times 10^{19}$ atoms/cm$^3$. In the 160 $^\circ$C sample, the nitrogen concentration drops off over a depth of $\sim$15 nm to below $1\times10^{19}\ \text{atoms}/\text{cm}^3$ making it irrelevant to the RF performance. The concentrations of C and O are surprisingly significant: $\sim$4.5$\times10^{20}$ atoms/cm$^3$ at a depth of 50 nm. This is much higher than that of the high-temperature nitrogen-doped sample and is consistent with RF measurements.  

There are several possible sources for the C and O. The high purity nitrogen gas used during the 160 $^\circ$C bake contained $\leq 5$ ppm O$_2$, $\leq 3$ ppm H$_2$O, and an unknown amount of C sources such as CO, CO$_2$, and hydrocarbons (i.e. $< 1$ ppm). However, the continuous flow of gas provided the cavity and sample with a constant source of C and O. This can account for the O content of the sample but cannot fully explain the C content. Another source of C could be from pre-existing surface carbon or carbides and back-flow from the furnace pumps.

\section{Diffusion Model}
To create a simple model for the diffusion of C, O, and N in bulk Nb, we solved Fick's second law,
\begin{equation}
\label{eq:one}
\frac{\partial c}{\partial t} = D \frac{\partial^2 c}{\partial x^2}.
\end{equation}
The relevant parameters of Eq.\ \ref{eq:one} are the concentration of the impurity, $c$, as a function of depth into the niobium, $x$, diffusion time, $t$, and the diffusion coefficient, $D$.  The solution to Fick's law with the appropriate boundary conditions is
\begin{equation}
c(x,t) =C'+(C''-C') \text{erfc}\left(\frac{x}{2\sqrt{Dt}}\right).
\end{equation}
$C'$ and $C''$ are the initial concentration of the impurity throughout the niobium and the impurity concentration at the surface, respectively. The model, calculated numerically, takes into account the temperature dependence of the diffusion coefficient with the exception of the cool down period from 800 to 160 $^\circ$C. The coefficients obtained by fitting the model to the  SIMS data are summarized in Table \ref{tab:one}. The measured diffusion curves plotted along with the model are shown in Fig.\ \ref{fig:one}. 
\begin{table}
\centering
\caption{Diffusion coefficients for C, O, and N in bulk Nb.}
\label{tab:one}
\def\arraystretch{1.5}
\begin{tabular}{c | c | c}
	Element&  $T$ [$^\circ$C] &  $D$ [cm$^2$/s]
	\\ \hline
	\multirow{2}{*}{C}	& 160 & $5.12 \times 10^{-17}$	\\[0.2em]
				& 800 & $1.73 \times 10^{-13}$ 	\\[0.2em]
	\hline
	\multirow{2}{*}{N}	& 160 & $3.56 \times 10^{-17}$	\\[0.2em]
				& 800 & $1.11 \times 10^{-13}$	\\[0.2em]
	\hline
	\multirow{2}{*}{O}	& 160 & $2.50 \times 10^{-18}$	\\[0.2em]
				& 800 & $6.89 \times 10^{-14}$	\\[0.2em]
	\hline
\end{tabular}
\end{table}

Tthe carbon concentration profile has two distinct regions. The region with a steeper slope ($x < 200$ nm) can be accounted for by the diffusion of carbon impurities in the doping atmosphere during the 160 $^\circ$C bake. The extended region ($x > 200$ nm) can be explained by the diffusion of surface carbon during the initial 800 $^\circ$C vacuum bake.

\section{Material Properties}
At $\ell = 7$ nm, the penetration depth of niobium for $T \ll T_c$ is $\sim$100 nm.  We will take the concentration measured by SIMS of C and O at a depth of 50 nm to to calculate an approximate value for $\ell$. At this depth, the concentration of C and O is $\approx 4.4\times10^{20}$ atoms/cm$^3$. This corresponds to $\approx$ 0.8 at. \% C, O. The change in resistivity can be calculated for these concentrations using the formula\cite{H08}:

\begin{equation}
\Delta \rho = a \cdot c'. \\[2ex]
\end{equation}

Here $a$ is a constant that is $4.3\times10^{-8}\ \Omega\cdot$m for carbon and $4.5\times10^{-8}\ \Omega\cdot$m for oxygen\cite{S81}. This yields $\Delta \rho_{\text{C}} = 3.4\times10^{-8}\ \Omega\cdot$m and $\Delta \rho_{\text{O}} = 3.6\times10^{-8}\ \Omega\cdot$m. The mean free path is related to the change in resistivity by\cite{G16}:

\begin{equation}
\ell = \frac{\sigma}{\Delta \rho_{\text{C}}+\Delta \rho_{\text{O}}}.
\end{equation}

\begin{figure}[h]
\centering{
\includegraphics[width=0.90\columnwidth]{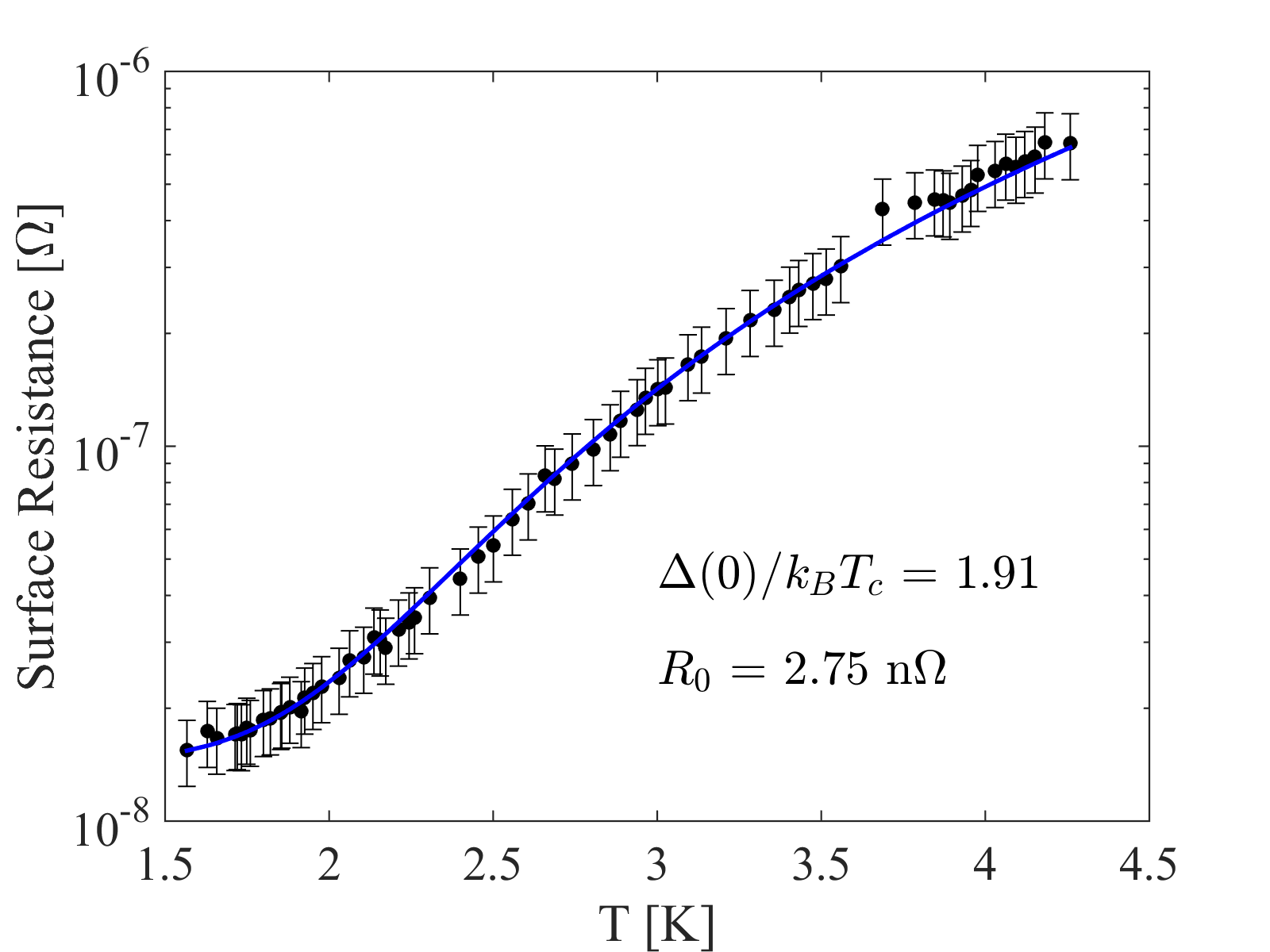}
\includegraphics[width=0.90\columnwidth]{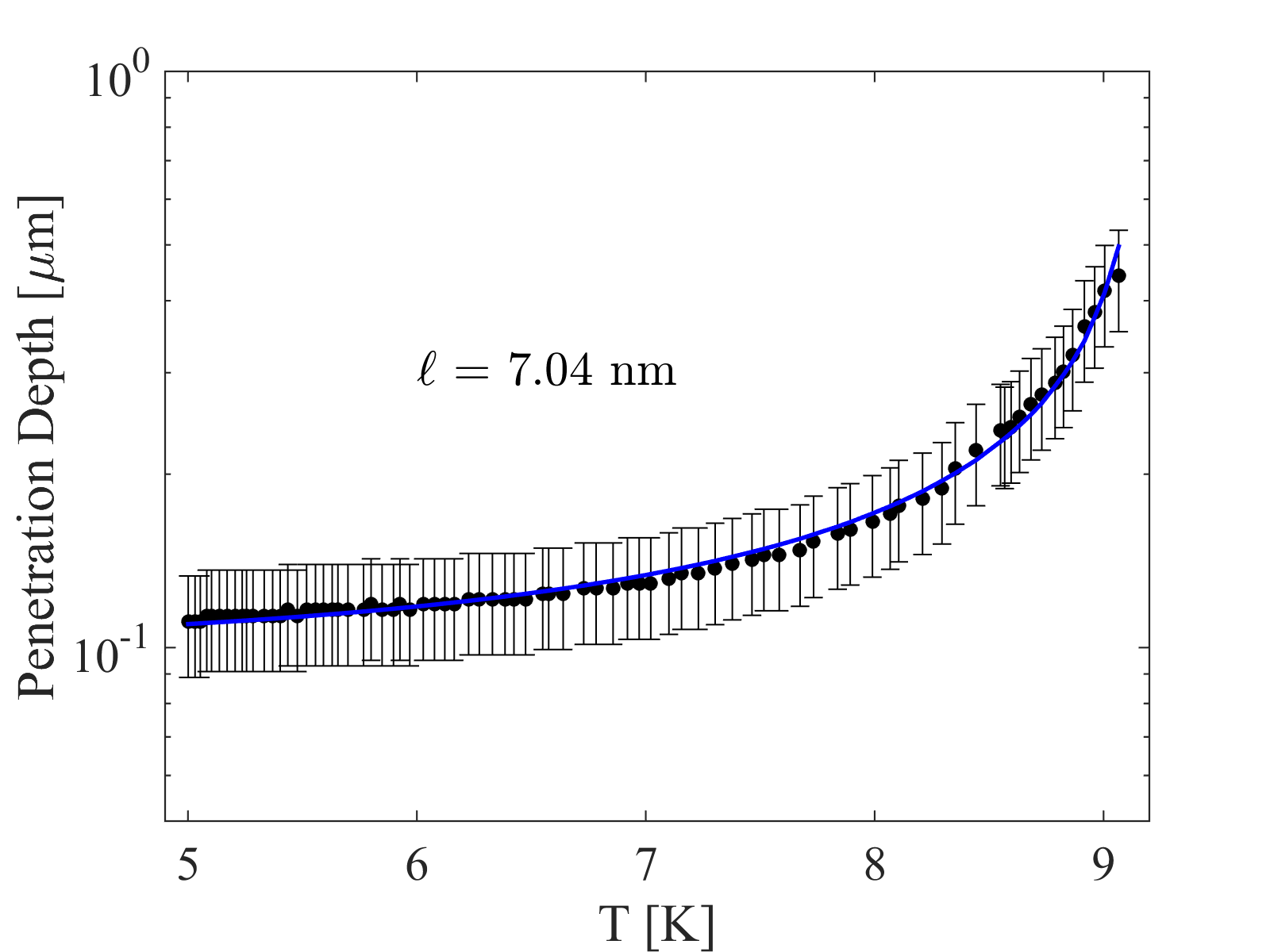}
\caption{\textbf{Top:} Measured RF surface resistance, $R_S$, as a function of $T$ and the BCS fit (solid line) with the extracted residual resistance, $R_0$, and quasi-particle energy gap, $\Delta(0)/k_BT_c$. \textbf{Bottom:} Penetration depth as a function of $T$ and the extracted mean free path $\ell$ with BCS fit (solid line).}
\label{fig:two}}
\end{figure}

From Goodman and Kuhn\cite{GK68}, $\sigma = 0.37\times10^{-15}\ \Omega\cdot$m$^2$ yielding a mean free path estimate of $\ell \approx 5$ nm. This is in agreement with the mean free path extracted from RF data (see Fig.\ \ref{fig:two}). At a depth of 3 nm, the nitrogen concentration is at its maximum of 0.12 at.\ \% N. The resulting change in resistivity is $6.0 \times 10^{-9}\ \Omega\cdot$m where, in this case, $ a = 5.2\times10^{-8}\ \Omega\cdot$m\cite{S81}. This leads to a mean free path estimate of $\ell \approx 60$ nm. However, this drops off sharply with depth. At a depth of 50 nm the concentration of nitrogen is 0.01 at. \% corresponding to a mean free path of $\sim$712 nm. This further confirms the role that carbon and oxygen play in the lowering of the mean free path in the RF penetration layer, ruling out nitrogen as the cause.

\section{Conclusion}
We have shown that treating niobium in a continuously flowing `nitrogen atmosphere' with trace impurities at 160 $^\circ$C introduces interstitial carbon and oxygen in concentrations similar to that of high-temperature nitrogen-doped cavities leading to a reduction of the electronic mean free path and subsequent increase in quality factor with increasing accelerating fields. Nitrogen does not play any significant role in the lowering of the mean free path nor in the effects on the quality factor in these low temperature treatments. This introduces the advantage of the elimination of post-treatment electro-polishing that is required for nitrogen-doped cavities. Further investigations into the effects of oxygen and carbon-doping of niobium are forthcoming. \\

\bibliography{OxyCarbonImpurities}

\end{document}